\documentclass[a4paper,12pt]{article}
\usepackage{amsmath}
\usepackage{graphicx}
\title{\textbf{Analytical Analysis and Numerical Solution\\
of Two Flavours Skyrmion\footnote{\textit{This topics was presented at WCS2K5 UGM, Yogyakarta, Indonesia, 8 September 2005.}}}}
\author{\textbf{Miftachul Hadi}$^1$, \textbf{Irwandi Nurdin}$^2$, \textbf{Denny Hermawanto}$^3$\\[0.5cm]
$^1$Applied Mathematics for Biophysics Group\\
Physics Research Centre, Indonesian Institute of Sciences (LIPI)\\
Puspiptek, Serpong, Tangerang 15314, Banten, Indonesia\\
E-mail: miftachul.hadi@lipi.go.id \\[0.5cm]
$^2$Department of Physics, Faculty of Mathematics and Natural Sciences\\
University of Syiah Kuala, Banda Aceh, Indonesia\\[0.5cm]
$^3$Research Centre for Calibration, Instrumentation and Metrology\\
Indonesian Institute of Sciences (LIPI)\\
Puspiptek, Serpong, Tangerang 15314, Banten, Indonesia
{}}
\date{}
\begin{document}
\maketitle
\begin{abstract}
Two flavours Skyrmion will be analyzed analytically, in case of static and rotational Skyrme equations. Numerical solution of a nonlinear scalar field equation, i.e. the Skyrme equation, will be worked with finite difference method. This article is a more comprehensive version of \textit{SU(2) Skyrme Model for Hadron} which have been published at Journal of Theoretical and Computational Studies, Volume \textbf{3} (2004) 0407.
\end{abstract}

\section{Introduction}
\textit{Nonlinear physics} is a general phenomenon in physics. It includes very long range from e.g. particle, nuclear, condensed matter, fluids, plasmas, biophysics (for example: nonlinear diffusion-reaction, DNA), to cosmology. 

\textit{Soliton} is defined as classical solution of nonlinear wave equation which has properties: finite total energy, localized, nondispersive, stable with its profile of energy density distribution is like pulse centered in finite space \cite{mif04}. 

\textit{Skyrmion} is a \textit{topological soliton} in three dimensions. Following, we will discuss soliton as \textit{baryon model}, especially \textit{Skyrme model} for hadron which is formulated in \textit{extended nonlinear Sigma model} \cite{mif04}, \cite{hans05}.

\section{Topological Solitons}
Suppose that $F_\phi=0$ is a differential equation or a system of differential equation involving a field or a set of fields denoted by $\phi$ which are functions of $D$ space coordinates $\vec{x}$ and a time coordinate $t$. We shall assume that $F_{\phi}=0$ is the equation of motion which results from some Lagrangian filed theory. We also assume that the system has a corresponding "energy" $\varepsilon_\phi$ and "energy density" $E_\phi(\vec{x},t)$ \cite{bal91}
\begin{equation}\label{1}
\varepsilon_\phi=\int d^Dx~E_\phi(\vec{x},t)
\end{equation}
where for all allowed field configurations $\phi$, $E_\phi$ is greater than or equal to zero. If $E_{\phi}$ is zero for all $\vec{x}$, we call $\phi$ \textit{a ground state} or \textit{a vacuum solution}, and denote it by $\phi_{vac}$. The vacuum solution need not be unique.

Now suppose that $\phi=\phi_{cl}$ is \textit{non-vacuum solution} to $F_\phi=0$. Following Coleman, we shall call $\phi_{cl}(\vec{x},t)$ a \textit{soliton solution} if the following properties are satisfied \cite{bal91}:
\begin{itemize}
\item[(1)]
$\varepsilon_{\phi=\phi_{cl}}$ is \textit{finite}.
\item[(2)]
$E_{\phi=\phi_{cl}}(\vec{x},t)$ is \textit{nonsingular (finite)} for all value of $\vec{x}$ and $t$, and \textit{localized} for all times $t$. A solution is localized at any time $t$ if there is a bounded region of space defined by $E_{\phi=\phi_{cl}}(\vec{x},t)\geq\delta$, $\delta$ being any arbitrary number, fulfilling
\begin{equation}\label{2}
0<\delta<\text{max}_{\vec{x}}~E_{\phi=\phi_{cl}}(\vec{x},t).
\end{equation}
We say that the solution is localized for all times $t$ if the bounded region can be chosen independent of $t$.
\item[(3)]
$\phi_{cl}(\vec{x},t)$ is \textit{nonsingular}.
\item[(4)]
$\phi_{cl}(\vec{x},t)$ is \textit{nondissipative}.\\
Concerning (4), a solution $\phi_{cl}(\vec{x},t)$ is considered dissipative if
\begin{equation}\label{3}
\text{lim}_{t\rightarrow\infty}\text{max}_{\vec{x}}~E_{\phi=\phi_{cl}}(\vec{x},t)=0.
\end{equation}
In our discussions, we will be concerned primarily with "static" solutions for which we denote $\phi_{cl}(\vec{x},t)$ by $\Phi_{cl}(\vec{x},t)$. 

\textit{If we exclude the vacuum, then static, nonsingular, localized solutions are automatically nondissipative}. Thus for static solitons, we only require (1), (2) and (3). 

We will, however, be interested in static solutions which satisfy yet another requirements; namely that they be
\item[(5)]
\textit{classically stable}.
We call a static solution "classically stable" if under all possible infinitesimal fluctuations $\delta\phi$ of the fields $\phi$ about the static solution, 
\begin{equation}\label{4}
\varepsilon_{\phi_{cl}+\delta\phi}\geq\varepsilon_{\phi_{cl}}.
\end{equation}
\end{itemize}
Variational modes $\delta_0\phi$ which leave the energy unchanged are said to be "zero" frequency modes" or "zero modes". Thus, for zero modes, $\varepsilon_{\phi_{cl}+\delta_{0}\phi}=\varepsilon_{\phi_{cl}}$. \textit{Zero modes are generally associated with various symmetries of the soliton}. Excluding the zero modes, a classically stable solution is a local minimum of the energy.

Although the question of \textit{stability} is ultimately a dynamical one, \textit{topology} often plays an important role. We shall now elaborate on this point. Let $Q$ be \textit{the set} of all finite energy and nonsingular field configurations $\phi$ at some fixed time $t$. It is thus \textit{the configuration space} of the system. \textit{A subset} $Q_1$ of $Q$ is said to be \textit{(path-)connected} if any field $\phi_1$ of $Q_1$ can be \textit{continuously deformed} to any other field $\phi_1'$ in $Q_1$. 

Two subsets $Q_1$ and $Q_2$ of $Q$ are said to be \textit{disconnected} if any field $\phi_1$ of $Q_1$ cannot be continuously deformed to any field $\phi_2$ in $Q_2$. We shall consider the case where $Q$ has $N>1$ disconnected components $Q_n$, each $Q_n$ being connected. $Q$ is the union of all these disconnected components,
\begin{equation}\label{5}
Q=\cup^N_{n=1}Q_n.
\end{equation}

Let $\phi(\vec{x})$ and $\phi'(\vec{x})$ be two \textit{fields} in $Q$. $\phi(\vec{x})$ and $\phi'(\vec{x})$ are \textit{homotopic} to each other, and write $\phi\sim \phi'$, if there exists a sequence of fields $\phi^{(\tau)}(\vec{x})$, $0\leq\tau\leq 1$, which is continuous in $\tau$ and $\vec{x}$, with $\phi^{(0)}(\vec{x})=\phi(\vec{x})$ and $\phi^{(1)}(\vec{x})=\phi'(\vec{x})$. With this definition, all fields within one component $Q_n$ are "homotopic" to each other, while a field $\phi^{(n)}$ belonging to $Q_n$ is not homotopic to a field $\phi^{(n')}$ belonging to $Q_{n'}$ when $n'\neq n$ \cite{bal91}.

We can treat all our examples assuming that $n$ is countable. We make this assumption: \textit{What is the physical significance of this classification?} Consider the initial conditions $(\phi^{(n)},d\phi^{(n)}/dt)$ at time $t=0$ for the equations of motion. (These equations for simplicity are assumed to be second order in time). Assume that $\phi^{(n)}\in Q_n$. After a lapse of time $T$, suppose that $\phi^{(n)}$ becomes $\phi'^{(n)}$ and $d\phi^{(n)}/dt$ becomes $d\phi'^{(n)}/dt$. Since time evolution is assumed to be a continuous operation, it follows that $\phi'^{(n)}$ is homotopic to $\phi^{(n)}$. Hence, $\phi'^{(n)}\in Q_n$. In other words, the value of $n$ associated with the field $\phi^{(n)}$ is a constant of the motion. Hence, if we define  $Q_0$ to be the sector which contains the vacuum solution $\phi_{vac}$, then the configurations $\phi^{(n)}\in Q_n,~n\neq 0$, cannot be time evolved to $\phi_{vac}$ or in fact, to any other $\phi^{(m)}\in Q_m$ with $m\neq n$. For such reasons, the sectors $Q_{n\neq0}$ of field configurations are said to be \textit{topological stable}. Soliton configurations which fulfill the five properties listed previously and which are in topologically stable sectors are called \textit{topological solitons} \cite{bal91}. 

\section{Two Flavours Skyrmion as Hadron}
In particle physics, \textit{flavour} is a quantum number of elementary particles. In quantum chromodynamics (QCD), flavour is a global symmetry. In the electroweak theory, on the other hand, this symmetry is broken, and flavour changing processes exist \cite{wiki-flav}.

If there are two or more particles which have identical interactions, then they may be interchanged without affecting the physics. Any (complex) linear combination of these two particles give the same physics, as long as they are orthogonal or perpendicular to each other. In other words, the theory possesses symmetry transformations such as $M\left({u\atop d}\right)$, where $u$ and $d$ are the two fields, and $M$ is any $2\times 2$ unitary matrix with a unit determinant. Such matrices form a Lie group called $SU(2)$, \textit{special unitary group}. This is an example of flavour symmetry. The term "flavour" was first coined for use in the quark model of hadrons in 1968 \cite{wiki-flav}.

In order to understand the charge radius of nucleon which have size roughly 1 Fermi, Tony Hilton Royle Skyrme in 1962 proposed idea that strongly interacting particles (hadrons) were locally concentrated static solution of extended nonlinear Sigma (Chiral) model. In (3+1)-dimensions of space-time, we will observe Skyrme model which describes hadrons as solitons (Skyrmions) from nonlinear Sigma (chiral) model field theory in internal symmetrical group of SU(2): two flavours Skyrmion. 

\textit{Skyrme's idea is unifying bosons and fermions in a common framework which provide a fundamental 
fields model consisted of only pion. The nucleon was obtained, as a certain classical configuration of the pion fields. Skyrmion is a topological soliton object, i.e. solution to the classical field equation with localized energy density. Various atoms should correspond to vortices of different connectivities in some underlying liquid}.

Following, some ideas which are proposed by Skyrme in relations with baryon model \cite{mif04}, \cite{val92}, \cite{mift04}:
\begin{itemize}
\item[(1)] Meson field can take its value in $S_3$ manifold. As the result of this assumption, Skyrme discovers conserved quatity i.e. \textit{topological charge} or \textit{winding number} and he interpretates it as \textit{baryon number}.
\item[(2)] Solution of Skyrme field equation in spherical coordinate takes form: 
\begin{equation}\label{6}
F_a (r,\theta,\phi) = g(r)n_a (\theta,\phi)
\end{equation}
where $g(r)$ is profile function which has spherical symmetric character and $n_a$ ($a=1,2,3$) is component of unit vector, 
$\boldsymbol{\widehat n}$. $F_a (r,\theta,\phi)$ is called \textit{Skyrme ansatz} or \textit{ hedgehog}.
\item[(3)] Skyrme ansatz describes \textit{stable extended particle} with unit topological charge.
\item[(4)] Skyrme ansatz with unit topological charge \textit{can be quantized} as fermion, so it is possible to identify state of isotopic spin, $I$, and total spin, $J$, with nucleon doublet in case of  $I=J=\frac{1}{2}$, and with $\Delta$ resonance in case of $I=J=\frac{3}{2}$.
\end{itemize}

\section{SU(2) Skyrme Model}
SU(2) Skyrme model is very simple model, because it only consists of two flavours. This model is described by $U=U(x^\mu)$ function which has SU(2) group valued of (3+1)D space-time coordinates \cite{mif04,mift04,hans02}.

The dynamics is determined by \textit{action}
\begin{equation}\label{7}
S=\int d^4x~\mathcal{L},
\end{equation}
where
\begin{equation}\label{8}
\mathcal{L}=\text{Tr}\left[-\frac{F^2}{16}L_\mu L^\mu+\frac{1}{32a^2}\left[L_\mu,L_\nu\right]\left[L^\mu ,L^\nu\right]+\frac{F^2}{16}
M_\pi^2(U^{-1}+U-2I)\right]
\end{equation}
is the related \textit{Lagrangian density} and
\begin{equation}\label{9}
L_\mu=U^{-1}\partial_\mu U
\end{equation}
are \textit{left chiral currents}, where $\mathit{F}\cong{123}$ MeV is pion decay constant and $a$ is dimensionless constant. \textit{The first term in equation (\ref{8}) is the SU(2) chiral model Lagrangian density, the second term is Skyrme term for stabilizing solitonic solution. The last term descibes the mass term where $\mathit{M}_\pi$ is pion (meson) mass}.

Euler-Lagrange equation of SU(2) Skyrme model is derived from \textit{least action principle} 
\begin{equation}\label{10}
\delta S = 0.
\end{equation}
If we take \textit{variation of action} to equation (\ref{7}), we obtain
\begin{equation}\label{11}
\delta S=\int d^4x~\delta\mathcal{L}.
\end{equation}
Substitute equation (\ref{8}) into (\ref{11}), and use least action principle (\ref{10}) \cite{mif04}
\begin{equation}\label{12}
\partial_\mu\left({L^\mu-\frac{1}{{a^2F^2}}\left[{L_\nu,\left[{L^\mu,L^\nu}\right]}\right]} \right)+\frac{1}{2}M_\pi^2(U-U^{-1})=0.
\end{equation}

In general case (nonstatic), energy of SU(2) Skyrme model is
\begin{equation}\label{13}
E = \int {d^3 } x~T^{00},
\end{equation}
where $T^{00}$ is \textit{energy-momentum tensor}. Explicitly, 
\begin{eqnarray*}
E
&=&\int d^3x~\text{Tr}\left[-\frac{F^2}{16}L_a
L_a-\frac{1}{32a^2}[L_a,L_c][L_a,L_c]\right.\nonumber\\&&
\left.
+\frac{F^2}{16}L_o L_o+\frac{1}{16a^2}[L_o,L_a][L_o,L_a]-\frac{F^2}{16}
M_\pi^2(U^{-1}+U-2I)-\frac{F^2}{8}L_o L_o\right.\nonumber\\&&
\left.-\frac{1}{8a^2}[L_a,L_o][L_a,L_o]\right]\\
&=&E_{\text{static}}+E_{\text{rotation}}
\end{eqnarray*}\begin{equation}\end{equation}
where
\begin{equation}\label{15}
E_{\text{static}}=-\int{d^3x~\text{Tr}\left[{\frac{{F^2}}{{16}}L_a^2+\frac{1}{{32a^2}}\left[{L_a,L_c}\right]^2 +\frac{{F^2}}{{16}}M_\pi^2(U^{-1}+U-2I)}\right]}
\end{equation}
and
\begin{equation}\label{16}
E_{\text{rotation}}=-\int{d^3x~\text{Tr}\left[{\frac{{F^2 }}{{16}}L_0^2+\frac{1}{{16a^2}}\left[{L_0,L_a} \right]^2}\right]}.
\end{equation}

Solitonic properties of SU(2) Skyrme model for static energy in equation (\ref{15}) is studied by scaling the spatial coordinates 
\begin{equation*}
x\rightarrow2\widetilde{x}/aF 
\end{equation*}
and express energy in $F/4a$, i.e. by taking 
\begin{equation*}
(F/4a)=(1/12\pi^2). 
\end{equation*}
In this unit, equation (\ref{15}) becomes
\begin{equation}\label{17}
E_{\text{static}}=\frac{1}{{12\pi^2}}\int{d^3}x\left(-\frac{1}{2}\right)\text{Tr}\left[{L_a^2+\frac{1}{8}\left({\left[{L_a,L_c}\right]^2+m_\pi^2(U^{-1}+U-2I)}\right)}\right]
\end{equation}
where
\begin{equation}\label{18}
m_\pi=2M_\pi/aF.
\end{equation}
Euler-Lagrange equation (\ref{12}) in static case becomes 
\begin{equation}\label{19}
\partial_a\left({L_a-\frac{1}{4}\left[{L_c,\left[{L_a,L_c}\right]}
\right]}\right)-\frac{{m_\pi^2}}{2}(U-U^{-1})=0.
\end{equation}

\section{Scale Stability}
Let us look at scale transformation below
\begin{equation}\label{20}
x\rightarrow \lambda x.
\end{equation}
We find that $\mathit L_a$ currents, by scale transformation (\ref{20}), transform into
\begin{equation}\label{21}
L_a(x)\to U^{-1}(\lambda x)\frac{\partial U(\lambda x)}{\partial x^a} =
\lambda L_a (\lambda x).
\end{equation}
The effect of scale transformation to static energy (\ref{17}), by ignoring pion mass term (because pion mass is small), is
\begin{equation}\label{22}
E\left[\lambda\right]_{\text{static}}=\frac{1}{\lambda}E_\sigma+\lambda E_{\text{Sky}}
\end{equation}
where $E_\sigma$ is \textit{static chiral energy term} and $E_{\text{Sky}}$ is \textit{Skyrme energy term}.

From equation (\ref{22}), we obtain
\begin{equation}\label{23}
\left.\frac{dE\left[\lambda\right]}{d\lambda}\right|_{\lambda=1}
=\left.\left(-\frac{1}{\lambda^2}E_\sigma+E_{\text{Sky}}\right)\right|_{\lambda=1}
=-E_\sigma+E_{\text{Sky}},
\end{equation}
and
\begin{equation}\label{24}
\left.\frac{d^2 E\left[\lambda\right]}{d\lambda^2}\right|_{\lambda=1}
=\frac{2}{\lambda^3}\left.{E_\sigma}\right|_{\lambda=1}=2E_\sigma.
\end{equation}
The requirement for \textit{extremum condition} is
\begin{equation}\label{25}
\frac{{dE\left[ \lambda  \right]}}{{d\lambda }} = 0.
\end{equation}
We apply extremum condition, (\ref{25}), to equation (\ref{23}), we obtain
\begin{equation}\label{26}
E_\sigma=E_{Sky},
\end{equation}
which it shows 
\begin{equation}\label{27}
E_{\sigma}\geq 0.
\end{equation}
So that equation (\ref{24}) fulfills condition
\begin{equation}\label{28}
\frac{{d^2E\left[\lambda\right]}}{{d\lambda^2}}> 0.
\end{equation}
Equation (\ref{28}) is \textit{minimum stable condition} which implies that \textit{static energy (\ref{22}) is stable against scale perturbation}.

\section{Topological Charge}
Static energy of SU(2) Skyrme model can be expressed, by ignoring pion mass, as 
\begin{equation}\label{29}
E_{\text{static}}=\frac{1}{{12\pi^2}}\int{d^3}x\left(-\frac{1}{2}\right)\text{Tr}\left[{\left({L_a\pm \frac{1}{4}\epsilon_{abc}\left[{L_b,L_c}\right]}\right)^2}\right]\mp\frac{1}{{24\pi^2}}\int {d^3}x~\epsilon_{abc}\text{Tr}~[L_aL_bL_c]
\end{equation}
where $\epsilon_{abc}$ is Levi-Civita symbol, $\epsilon_{abc}=\delta^{abc}_{123}$. 
At energy lower bound
\begin{equation}\label{30}
E_{\text{static}}\geq B,
\end{equation}
where
\begin{equation}\label{31}
B=-\frac{1}{{24\pi^2}}\int{d^3 x\epsilon _{abc}}\text{Tr}~(L_aL_bL_c).
\end{equation}
$B$ \textit{integral, (\ref{31}), is independent of space-metric tensor}, i.e. topological quantity which is known as \textit{topological charge} of SU(2) Skyrme model. 

\section{Static Skyrme Equation in Spherical Coordinate}
In spherical coordinate, $(r,\theta,\phi)$, static Skyrme equation has the following form \cite{mif04}
\begin{equation}\label{32}
\begin{split}
0
&=\partial_r\left(L_r-\frac{1}{4}\left\{\frac{1}{r^2}[L_\theta,[L_r,
L_\theta]]+\frac{1}{r^2\sin^2\theta}[L_\phi,[L_r,L_\phi]]\right\}\right)\\
&+\partial_\theta
\left(\frac{1}{r^2}L_\theta-\frac{1}{4}\left\{\frac{1}{r^2}[L_r,[L_\theta,L_r]]+\frac{1}{r^4\sin^2\theta}[L_\phi,[L_\theta,L_\phi]]\right\}\right)\\
&+\partial_\phi\left(\frac{1}{r^2\sin^2\theta}L_\phi-\frac{1}{4}\left\{\frac{1}{r^2\sin^2\theta}[L_r,[L_\phi,L_r]]+\frac{1}{r^4\sin^2\theta}[L_\theta,[L_\phi,L_\theta]]\right\}\right).
\end{split}
\end{equation}
Its solution takes form
\begin{equation}\label{33}
U(r)=\exp(iF_a(r,\theta,\phi)\sigma_a),
\end{equation}
where
\begin{equation}\label{34}
F_a(r,\theta,\phi)=g(r)n_a(\theta,\phi)
\end{equation}
is Skyrme ansatz, $g(r)$ is profile function, $n_a$, $a=1,2,3$ is component of unit vector, $\widehat{\boldsymbol{n}}$, in the internal space of SU(2), and $\sigma_a$ is Pauli matrix.

\section{Skyrmion Static Energy and Its Solution}
As a result of scaling spatial coordinates, by ignoring pion mass, static energy of SU(2) Skyrme model can be stated as \cite{mif04}
\begin{eqnarray}\label{35}
E_{\text{static}}=\frac{1}{12\pi^2}(4\pi)\int r^2 dr\left[\left(\frac{dg}{dr}\right)^2+\frac{2}{r^2
}\sin^2g\left(1+\left(\frac{dg}{dr}\right)^2\right)+\frac{1}{r^4}\sin^4g\right].
\end{eqnarray}
From equation (\ref{35}), by using least action principle
\begin{equation}\label{36}
\delta_g E_{static} = 0,
\end{equation}
we can derive Euler-Lagrange equation for profile function \cite{mif04}
\begin{equation}\label{37}
\frac{{d^2g}}{{dr^2}}\left[{1+\frac{2}{{r^2}}\sin^2g}\right]
+\left({\frac{{dg}}{{dr}}}\right)^2\left[{\frac{1}{{r^2}}\sin 2g}
\right]+ \left({\frac{{dg}}{{dr}}} \right)\left({\frac{2}{r}}\right)-\frac{1}{{r^2}}
\sin 2g-\frac{1}{{r^4}}\sin^2 g\sin 2g=0.
\end{equation}
Equation (\ref{37}) is second order of nonlinear differential equation. 

The solution of equation (\ref{37}) will be worked numerically with finite difference method. In order that static energy (\ref{35}) has finite value at $r=0$ and $r=\infty$, then profile function $g(r)$ must fulfills \textit{boundary conditions}
\begin{equation}\label{38}
g(0)=\pi,~~~ g(\infty)=0.
\end{equation}

\section{Quantized Rotational Energy}
Skyrmion quantization is worked by involving time dependent of $U(\boldsymbol{r})$ as 
\begin{equation}\label{41}
U(\boldsymbol{r})\to U(\boldsymbol{r},t)=A(t)U(\boldsymbol{r})A(t)^\dag,
\end{equation}
where
\[A(t)\in SU(2)_{\text{internal}},\]
\begin{equation}\label{42}
AA^\dag=A^\dag A=I
\end{equation}
$A$ is time dependent unitary matrix.

By using equation (\ref{41}), rotational energy can be stated as \cite{mif04}, \cite{mift04}
\begin{eqnarray}\label{43}
E_{\text{rotation}}
&=&-\left({\frac{{\pi F^2}}{6}\int\limits_0^\infty{dr~r^2}\sin^2g+\frac{{2\pi}}{{3a^2}}\int\limits_0^\infty{dr~r^2\sin^2g\left[{\left({\frac{{dg}}{{dr}}}\right)^2+\frac{1}{{r^2}}\sin^2g}\right]}}\right)\text{Tr}\left({R^{-1}\frac{{\partial R}}{{\partial t}}}\right)^2\nonumber\\
&=&\frac{1}{2}~I~\text{Tr}~\boldsymbol{\Omega}^2
\end{eqnarray}
where
\begin{equation}\label{44}
\text{Tr}~\boldsymbol{\Omega}^2=-\text{Tr}\left(R^{-1}\frac{\partial R}{\partial t}\right)^2
\end{equation}
with $\boldsymbol{\Omega}$ is angular velocity matrix of Skyrmion,
and
\begin{equation}\label{45}
I=2\left(\frac{\pi F^2}{6}\int^\infty_0dr~r^2\sin^2g+\frac{2\pi}{3a^2}\int^\infty_0 dr~r^2\sin^2g\left[\left(\frac{dg}{dr}\right)^2+\frac{1}{r^2}\sin^2g\right]\right)
\end{equation}
is Skyrmion moment of inertia. 

\section{Finite Difference Method}
Assume that $g(x)$ is a continuous function of one variable. The value of this function is given only for discrete value, and equidistant $x$ \cite{ves}:
\begin{equation}\label{49}
g_k\equiv g(x_k),
\end{equation}
where $x_k\equiv x_0+k\Delta x$. Let us define quantity 
\begin{equation}\label{50}
\Delta g_k\equiv g_{k+1}-g_k
\end{equation}
which is called \textit{forward difference} at point $x_k$. Apply (\ref{50}) for higher order of forward difference, we get
\begin{equation}\label{51}
\Delta^2g_k\equiv \Delta g_{k+1}-\Delta g_k=g_{k+2}-2g_{k+1}+g_k,
\end{equation}
\begin{equation}\label{52}
\Delta^3g_k\equiv \Delta^2 g_{k+1}-\Delta^2 g_k=g_{k+3}-3g_{k+2}+3g_{k+1}-g_k
\end{equation}
and so on. General form of forward difference is: 
\begin{equation}\label{53}
\Delta^rg_k\equiv\Sigma_{i=0}^r(-1)^i\begin{pmatrix}
r\\i\end{pmatrix}g_{k+r-i}.
\end{equation}

\textit{Backward difference} is defined as
\begin{equation}\label{54}
\nabla g_k\equiv g_k-g_{k-1}
\end{equation}
and higher order of backward difference is 
\begin{equation}\label{55}
\nabla^2g_k\equiv \nabla g_k-\nabla g_{k-1}=g_k-2g_{k-1}+g_{k-2}
\end{equation}
so on. General form of backward difference is 
\begin{equation}\label{56}
\nabla^rg_k\equiv\Sigma_{i=0}^r(-1)^i\begin{pmatrix}
r\\i\end{pmatrix}g_{k-r+i}.
\end{equation}
Following, it is defined \textit{central difference} which is symmetric with $x_k$:
\begin{equation}\label{57}
\delta g_k\equiv g_{k+{1/2}}-g_{k-{1/2}}
\end{equation}
and higher order of central difference is 
\begin{equation}\label{58}
\delta^2g_k=g_{k+1}-2g_k+g_{k-1}
\end{equation}
so on. General form of central difference is 
\begin{equation}\label{59}
\delta^r g_k\equiv\Sigma_{i=0}^r(-1)^i\begin{pmatrix}
r\\i\end{pmatrix}g_{k-r/2+i}.
\end{equation}

If central difference is observed, $k+1/2$ index is out of the value which is served by discrete value. In order to obtain the integer value, mean value operator, $\mu$, is used to operate with $g(x)$ function as \textit{central mean}
\begin{equation}\label{60}
\mu g=\frac{1}{2}\left[g_{k+1/2}+g_{k-1/2}\right]
\end{equation}
and higher order of central mean is 
\begin{equation}\label{61}
\mu^2g_k\equiv\frac{1}{2}\left[\mu g_{k+1/2}+\mu g_{k-1/2}\right]
\end{equation}
so on. \textit{Central mean of central difference} is defined as 
\begin{eqnarray}\label{62}
\mu\delta g_k&\equiv& \frac{1}{2}\left[\delta g_{k+1/2}+\delta g_{k-1/2}\right]\nonumber\\
&=&\frac{1}{2}\left[g_{k+1}-g_{k-1}\right].
\end{eqnarray}

\section{Numerical Solution of Skyrme Equation}
Skyrme equation is stated in finite difference form, i.e. by substituting equation (\ref{54}), (\ref{58}), into Skyrme equation (\ref{37}) then we obtain
\begin{equation}\label{63}
\frac{\delta^2g_k}{\Delta r^2}\left[1+\frac{2}{r^2}\sin^2g_k\right]+\left(\frac{\nabla g_k}{\Delta r}\right)^2\left[\frac{1}{r^2}\sin~2g_k\right]+\left(\frac{\nabla g_k}{\Delta r}\right)\left(\frac{2}{r}\right)-\frac{1}{r^2}\sin~2g_k-\frac{1}{r^4}\sin^2g_k~\sin~2g_k=0.
\end{equation}
By reposisition of equation (\ref{63}), we get
\begin{eqnarray}
\delta^2g_k&=&\frac{-\left(\frac{\nabla g_k}{\Delta r}\right)^2\left[\frac{1}{r^2}\sin~2g_k\right]-\left(\frac{\nabla g_k}{\Delta r}\right)\left(\frac{2}{r}\right)+\frac{1}{r^2}\sin~2g_k+\frac{1}{r^4}\sin^2g_k\sin~2g_k}{\left[1+\frac{2}{r^2}\sin^2g_k\right]}\Delta r^2\nonumber\\
&=&\frac{-(\nabla g_k)^2\sin~2g_k-(\nabla g_k)2(\Delta r)r+\sin~2g_k\Delta r^2+\frac{1}{r^2}\sin^2g_k\sin~2g_k\Delta r^2}{r^2+2\sin^2g_k}
\end{eqnarray}
\begin{eqnarray}\label{65}
g_{k+1}&=&\frac{-(\nabla g_k)^2\sin~2g_k-(\nabla g_k)2(\Delta r)r+\sin~2g_k\Delta r^2+\frac{1}{r^2}\sin^2g_k\sin~2g_k\Delta r^2}{r^2+2\sin^2g_k}\nonumber\\
&-&g_{k-1}+2g_k.
\end{eqnarray}
The problems are initial values which are needed to be defined, i.e.
\begin{equation}\label{65.1} 
g(r=0),~~~g(r=dr). 
\end{equation}
Boundary conditions, which are given by Skyrme, are 
\begin{equation}\label{65.2}
g(r=0)=\pi,~~~g(r\rightarrow\infty)=0.
\end{equation}

Apply the boundary conditions (\ref{65.2}) for initial values (\ref{65.1}). It is suitable for $g(r=0)$, and $g(r=dr)$ must be chosen, so that it is suitable for $g(r\rightarrow\infty)=0$. This method is known as \textit{shooting method} \cite{irwandi2006}. The ploting result, as shown in Figure 1: Profile function $\phi(r)$, can be compared with trial and error method, i.e. 
\begin{equation}\label{66}
g(x)=4 \arctan~e^{-r}.
\end{equation}
\begin{figure}
\begin{centering}
\input{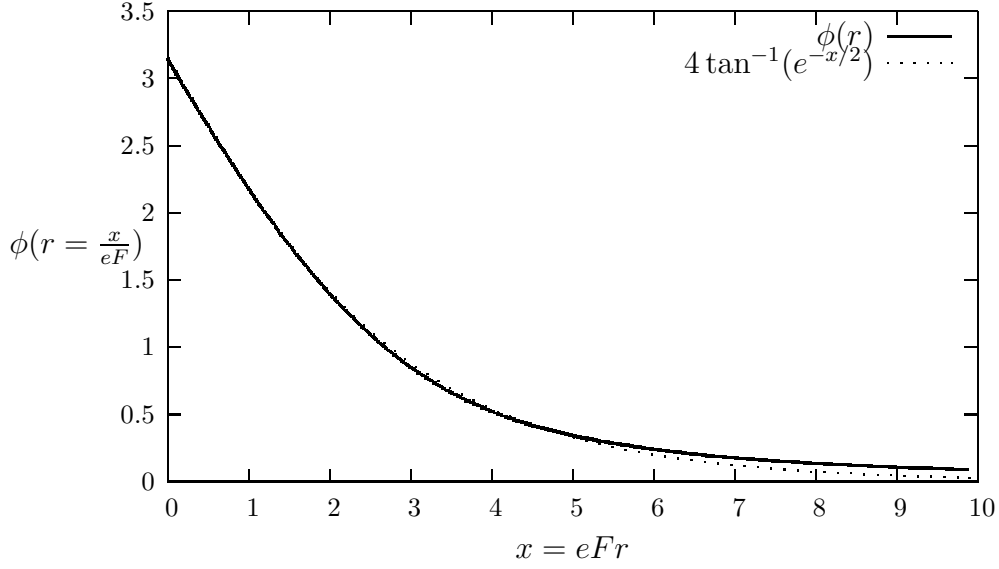}
\par\end{centering}
\caption{\label{cap:2-Fungsi-profile}Profile function $\phi(r)$}
\end{figure}
In order to calculate static energy (\ref{35}), the following integral is used
\begin{equation}\label{67}
E(g)=\frac{2\pi f_\pi}{e}\int_0^\infty dr\left[\left(r^2g'^2+2\sin^2g\right)+\sin^2g\left(2g'^2+\frac{\sin^2g}{r^2}\right)\right]
\end{equation}
where $g'=dg/dr$. Equation (\ref{67}), is solved numerically with \textit{trapezoid method} and it gives:
\begin{equation*}
23.2154~\pi f_\pi/e.
\end{equation*}
Let us calculate numerically $E_\text{static}$ (\ref{35}), using numerical value of $g(r)$
\begin{equation}\label{68}
E_{\text{static}}=\frac{1}{12\pi^2}(4\pi)\left(\frac{23.2}{2}\right).
\end{equation}
In the form of $F$ and $a$, using definition $(1/12\pi^2)=(F/4a)$, we obtain
\begin{equation}\label{69}
E_{\text{static}}=\frac{\pi F}{a}\left(\frac{23.2}{2}\right).
\end{equation}
The numerical value of $g(r)$ can be used to calculate
\begin{itemize}
\item[(1)] nucleon $(m_N)$ and delta static masses
$(m_\Delta)$;
\item[(2)] Skyrmion moment of inertia, $I$.
\end{itemize}

We get benefit from numerical value of profile function, $g(r)$, to calculate Skyrmion moment of inertia. It gives
\begin{equation}\label{46}
I=\frac{1}{4}\left(\frac{447}{Fa^3}\right).
\end{equation}
In quantum mechanics, angular momentum $\boldsymbol{J}$ is quantized as
\begin{equation}\label{47}
\boldsymbol{J}^2=j(j+1)\hbar^2
\end{equation}
where $j=0,\frac{1}{2},1,\frac{3}{2},2,\ldots$ 
\begin{equation*}
\hbar=\frac{h}{2\pi},
\end{equation*}
$h$ is Planck constant. Here, we use using natural unit, $\hbar=1=c$. 

Rotational energy is
\begin{equation}\label{48}
E_{\text{rotation}} = \frac{\boldsymbol{J}^2}{2I}=\frac{j(j+1)}{2I}.
\end{equation}

\section{Hadron Mass}
Skyrmion is solution of Skyrme equation which has finite energy. Based on Einstein energy-mass formula, Skyrmion mass is \begin{equation}\label{70}
m=\frac{E_{static}+E_{rotation}}{c^2}=M_{static}+M_{rotation}.
\end{equation}
From equation (\ref{48}) and (\ref{70}), we obtain
\begin{equation}\label{71}
m= M_{static} + \frac{{j(j + 1)}}{2I}.
\end{equation}
Wess-Zumino quantization condition \cite{bal91}, requires:
\[j = \frac{1}{2},\frac{3}{2},\frac{5}{2},\ldots\]
\textit{It means that Skyrmion is fermion}.

In case of nucleon with spin $\frac{1}{2}$,
$j=\frac{1}{2}$, equation (\ref{71}) gives
\begin{equation}\label{72}
m_N = M_{static} + \frac{\frac{1}{2}(\frac{1}{2}+1)}{2I}
\end{equation}
By substituting the values of $F = 123$ MeV and $a = 4.95$ into (\ref{69}) and (\ref{46}), give
\begin{equation}\label{73}
M_{static}\cong 905.08~
\text{MeV}
\end{equation}
and
\begin{equation}\label{74}
I\cong 7.49\times 10^{-3}.
\end{equation}
Substitute (\ref{73}) and (\ref{74}) into (\ref{72}) gives the result 
\begin{equation}\label{75}
m_N\cong 955.15~\text{MeV}.
\end{equation}

In case of delta particle with spin $\frac{3}{2}$, $j=\frac{3}{2}$,
equation (\ref{71}) gives
\begin{equation}\label{76}
m_\Delta=M_{static}+\frac{{\frac{3}{2}(\frac{3}{2}+1)}}{2I}.
\end{equation}
By substituting the values of $M_{static}$ and $I$ above, it gives
\begin{equation}\label{77}
m_\Delta\cong 1155.41~\text{MeV}.
\end{equation}

\section{Discussions}
The calculation of nucleon energy-mass gives the value about 955.15 MeV. The experimental value of nucleon energy-mass is about 939~MeV. The different value of nucleon energy-mass based on calculation and experiment is about 16.15 MeV.
The calculation of delta energy-mass gives the value about 1155.41 MeV. The experimental value of delta energy-mass is about 1232~MeV \cite{bal91}. The different value of delta energy-mass based on calculation and experiment is about 76.59 MeV.

The different values among calculation and experiment are becaused of SU(2) Skyrme model is very simple model, i.e. it consists of two flavours, $(u,d)$ or $(u,s)$ or $(d,s)$. It is necessary keeping in mind related with these different values \cite{hep54}:
\begin{itemize}
\item[(1)] Meson field Lagrangian only includes pseudoscalar field. Other low mass meson (vector) should be included. 
\item[(2)] In nature, there are three flavour families than two "light" flavour families which must be included in more realistic formula. 
\item[(3)] The effects of chiral and flavour symmetry breaking aren't calculated yet.
\item[(4)] $N_C$ (number of colour) correction of nucleon mass isn't included yet.
\end{itemize}
   
\section{Acknowledgment}
MH thank to Hans J. Wospakrik, Ph.D for great and eternal inspirations, Irwandi Nurdin and Denny Hermawanto for numerical solution. Toto Sudiro, Slamet for fruitful discussions. All kindly colleagues for their best supports. Physics Research Centre LIPI for research facilities, Andri, Ayu, Intan, Iim, Ike for their great hope. 

In depth, MH thank to beloved Mother for continuous praying and motivation. Special thank to beloved ones, Ika Nurlaila for huge support and patience, Aliya Syauqina Hadi for her funny and curiosity.

\section{Appendix: Flow Chart of Skyrme Profile Function Calculation and Its Integration}

\begin{verbatim}
clear;
clf;
okmainloop=0;
N=1;
dr=.001;
$%$dr=.01;

rf=50;
%rf=8;
rf=50;
rf=80;

nnl=length([0 dr:dr:rf rf+dr]);
g(nnl)=0;
n0=1;
n=2;
%%%%%%%%%
v=1+.0041
v0=v;
g(n0)=pi;%+0.01;
g(n)=g(n0)-v*dr;%0.02;

if okmainloop
% Looping utama program ------------
for r=[dr:dr:rf]
    n=n0+1;
    n1=n+1;

    gg=g(n);
    bg=g(n)-g(n0);
    bg2=bg*bg;
    sin2g=sin(gg)*sin(gg);
    sing2=sin(2*gg);
    r2=r*r;
    dr2=dr*dr;

    % x=eFr  (catto)
    g(n1)=(-2*r*bg*dr-sing2*(4*bg2-4*sin2g*dr2/r2-dr2))/(r2+8*sin2g)+2*g(n)-g(n0);
   
    n0=n0+1;
    %if (g(n1)<=0) break; end;
end
% akhir looping utama---------------- 
save simpan.mat
else
load simpan.mat    
end
 
rr=[0 dr:dr:rf rf+dr];
nn=n1;
gg=g(1:n1);
%subplot(2,1,1);
subplot(1,1,1);
plot(rr(1:nn-1),gg(1:nn-1),'b-');

rr=[0 dr:dr:rf rf+dr];
%rre=pi*exp(-rr);
rre=4*atan(exp(-rr));

hold on; 
plot(rr(1:nn-1),rre(1:nn-1),'r-');
grid on

%%%%%%%% negatif change to zero
%g=g.*(g>0);

% Statical calculation
IIEUc=0;
IIaUc=0;
IIbUc=0;
IICUc=0;
IIrUc=0;


jumlah=0;
dgdr2=v*v; % from initial value
gg=g(1);
sin2g=sin(gg)*sin(gg);


for n=[1:nnl-1]  
   dg=g(n+1)-g(n);
   %dg1=g(n+2)-g(n);
    if n==1 
        dgdr2=(dg*dg)/(dr*dr);
    else
        bg=g(n)-g(n-1);
        Dg=(g(n+1)-g(n-1))*.5;
        dgdr2=(dg*dg)/(dr*dr);
        %dgdr2=(dg*bg)/(dr*dr);
        %dgdr2=(Dg*Dg)/(dr*dr);
        
    end
    gg=g(n);
    r=n*dr;
    sin2g=sin(gg)*sin(gg);
    
    jumlah=jumlah+dr*gg;
    r2=r*r;
    IIEUc=IIEUc+ dr*( r2*dgdr2 + sin2g*(2+8*dgdr2+4*sin2g/(r*r)) );  % catto
    IIaUc=IIaUc+ dr*r2*sin2g*( 1 +  4*(dgdr2 + sin2g/(r*r))  );  % catto
    IIbUc=IIbUc+ dr*r2*(1-cos(gg))*(1 + dgdr2 + 2*sin2g/(r*r));
    IICUc=IICUc+ dr*r2*(1-cos(gg));
    IIrUc=IIrUc+ dr*r2*sin2g*(bg/dr);
    if (g(n+1)<0) break; end
end

IEUc=IIEUc*pi*4/8;
IaUc=IIaUc*pi*2/3;
IbUc=IIbUc*pi*1/2;
ICUc=IICUc*pi*1/2;
IrUc=IIrUc*2/pi;

save II.mat II* dr v0
\end{verbatim}

\end{document}